\documentclass[10pt]{iopart}

\usepackage{iopams}
\usepackage{color,hyperref,graphicx}
  \expandafter\let\csname equation*\endcsname\relax
  \expandafter\let\csname endequation*\endcsname\relax
\usepackage[pdftex,dvipsnames,usenames]{xcolor}
\usepackage{amsmath}
\usepackage{ulem}

\begin{document}

\title[Quantum teleportation through atmospheric channels]{Quantum teleportation through atmospheric channels}

\author{K Hofmann$^1$, A A Semenov$^{1,2,3}$, W Vogel$^1$, M Bohmann$^{1,4}$}
\address{$^1$ Arbeitsgruppe Theoretische Quantenoptik, Institut f\"ur Physik, Universit\"at Rostock, D-18051 Rostock, Germany}
\address{$^{2}$Institute of Physics, NAS of Ukraine, Prospect Nauky 46, UA-03028 Kiev, Ukraine}
\address{$^{3}$ Bogolyubov Institute for Theoretical Physics, NAS of Ukraine, Vul. Metrolohichna 14b, UA-03143 Kiev, Ukraine}
\address{$^{4}$ INO-CNR and LENS, Largo Enrico Fermi 2, I-50125 Firenze, Italy}
\ead{kevin.hofmann@uni-rostock.de}
\ead{martin.bohmann@ino.it}

\begin{abstract}
	We study the Kimble-Braunstein continuous-variable quantum teleportation with the quantum channel physically realized in the turbulent atmosphere.
	In this context, we examine the applicability of different strategies preserving the Gaussian entanglement [Bohmann et al., Phys. Rev. A \textbf{94}, 010302(R) (2016)] for improving the fidelity of the coherent-state teleportation.
	First, we demonstrate that increasing the squeezing parameter characterizing the entangled state is restricted by its optimal value, which we derive for realistic experimentally-verified examples.
	Further, we consider the technique of adaptive correlations of losses and show its performance for channels with large squeezing parameters.
	Finally, we investigate the efficiencies of postselection strategies in dependence on the stochastic properties of the channel transmittance.
\end{abstract}

% \pacs{}

\begin{indented}
	\item[]\today,\submitto{\PS}
\end{indented}

\vspace{2pc}

\maketitle
\ioptwocol

\section{Introduction}
	The no-cloning theorem \cite{Wootters1982,Dieks1982} is a fundamental concept of quantum physics, which states that it is impossible to copy any unknown quantum state without destroying it at the origin.
	Consequently, quantum states can only be transfered from one system to another.
	Such a process is usually referred to as quantum teleportation \cite{Bennett93,Pirandola15}, which is an essential building block of quantum networks.
	The original protocol \cite{Bennett93}, which has been proposed for teleporting states of two-level systems---qubits, has first been experimentally implemented by two groups in 1997  \cite{Bouwmeester1997,Boschi1998}.
	In a recent experiment, this protocol has been performed in the context of satellite mediated quantum communications for a ground-to-satellite channel over distances of up to 1400km \cite{Ren}.
	This experiment demonstrates the huge efforts made in order to establish teleportation channels for quantum information tasks on a global scale and underlines the general feasibility of such an endeavor.

	Continous-variable (CV) based quantum information processing~\cite{Braunstein2005} is an alternative approach, whose main idea consists in using systems with infinite-dimensional Hilbert spaces instead of qubits.
	Remote nodes of quantum networks can also be connected by using CV communication protocols.
	Quantum teleportation in the CV regime can be realised with the Braustein-Kimble (BK) protocol  \cite{Braunstein98}, which has been first experimentally implemented by Furusawa et al. \cite{Furusawa1998}.
	In this case, one uses Gaussian entanglement \cite{Horodecki2009,Simon2000,Adesso2014} shared between the communication parties as the main resource.
	Gaussian entanglement is also a recourse for other communication protocols such as CV entanglement swapping \cite{vanLoock1999}.

	Losses and other noise in optical channels may lead to the loss of Gaussian entanglement \cite{Filippov2014}.
	Particularly, Gaussian entangled states can be subdivided into classes with respect to their stability to constant losses \cite{Barbosa2011}.
	Particularly, a large class of states, whose entanglement can be verified with the Duan-Giedke-Cirac-Zoller (DGCZ) criterium \cite{Duan2000}, always preserves Gaussian entanglement under any constant loss conditions.
	An example of such DGCZ-states is the two-mode squeezed vacuum (TMSV) state, whose application has been proposed in the original version of the BK protocol.
	Consequently, the CV quantum teleportation can be performed, in principle, for entanglement shared through lossy channels \cite{Chizhov02}.
	However, the teleportation quality in the presence of large losses can be very low.

	In many situations, free-space channels have some practical advantages in comparison with optical fibers due to their mobility, possibility of satellite-mediated communications, etc.
	The establishment of such links could facilitate global quantum networks, which eventually may lead to a quantum Internet \cite{Kimble08}.
	Atmospheric links for quantum light were first tested in ground-to-ground experiments \cite{Ursin, Scheidl, Fedrizzi2009, Capraro, Yin, Ma, Peuntinger}.
	Further implementations could demonstrate the feasibility of satellite-mediated quantum communication using small-scale experiments \cite{Bourgoin13, Nauerth, Wang, Bourgoin15}.
	Ultimately, experiments with satellites have been reported \cite{Vallone15,Dequal16,Vallone16,Carrasco-Casado,Takenaka,Liao17b,Yin17a,Guenthner,Ren,Yin17b}, which include the successful implementation of satellite-based quantum key distribution \cite{Liao17b,Yin17b} and quantum state teleportation \cite{Ren}.

	Atmospheric free-space channels differ drastically from constant loss optical fiber links.
 	Due to atmospheric turbulent flows---causing temporal and spatial fluctuations of the optical properties of the atmosphere---the transmittance through such links varies in a random fashion.
 	This effect can be described by an appropriate probability distribution of the transmittance (PDT) \cite{Semenov2009}.
	Depending on the channel characteristics, advanced PDT models have been introduced \cite{beamwandering,VSV2016,Vasylyev17,Vasylyev18}, which accurately describe experiments \cite{Usenko,VSV2016,Vasylyev17} and can account for different weather conditions \cite{Vasylyev17}.
	Note that these atmospheric channel models can also be simulated in in-lab experiments; see, e.g., \cite{Bourgoin15b,Bohmann17b}.
	Such simulations can be an important tool for finding and testing quantum information applications for realistic free-space channels.

	The distribution of Gaussian entanglement through atmospheric channels has been rigorously studied in \cite{BSSV16}.
	Some conclusions of this work are of importance for the present consideration of CV teleportation through the atmosphere.
	First, the best result for the entanglement sharing can be reached if the entangled state does not have coherent displacements---this is exactly the scenario, which is used in the original BK protocol.
	Second, high values of squeezing of the TMSV states are in general not useful for such channels as they can lead to the total loss of Gaussian entanglement after the propagation through the atmosphere.
	There exists a maximal value of the squeezing parameter beyond which Gaussian entanglement distributed through the atmosphere vanishes.
	Third, for DGCZ-entangled states there always exists a possibility to preserve Gaussian entanglement by applying an adaptive channel correlation, at the expense of additional losses.
	Furthermore, post-selection scenarios can improve the entanglement transfer.
	Similar results were also reported for non-Gaussian states~\cite{Bohmann17a}.

	A general analysis of CV quantum teleportation protocols under the influence of fluctuating atmospheric losses is missing yet.
	In the literature, the influence of constant losses has been studied \cite{Chizhov02,Fiurasek02,Faria16}.
	Furthermore, a specific scenario of CV teleportation through a free-space channel has been considered in \cite{Zhang17}.
	This consideration is, however, limited to a particular loss scenario (beam-wandering PDT model) and focuses on the technical issue of the influence of the size of the receiver aperture.
	In particular, it does not provide general conclusions for CV quantum state teleportation through atmospheric channels.

	In this paper, we study the influence of atmospheric fluctuating-loss channels on the BK-CV quantum state teleportation protocol and propose strategies to improve the teleportation performance in such channels.
	In particular, we use the knowledge about Gaussian entanglement suffering from free-space losses and realistic free-space models in order to analyze and optimize CV teleportation for free-space applications.
	For this purpose we consider a standard task---the teleportation of unknown coherent states.
	We start by recalling the BK-CV teleportation protocol and the influence of constant losses on the teleportation.
	Next, we explain the model of atmospheric losses and show how it affects the teleportation process.
	Based on this consideration, we discuss different ways for an improved teleporation performance.
	This includes adaptive correlated channel losses and post-selection strategies.
	We also consider the case in which both modes of the Gaussian entangled state are transmitted through atmospheric channels.

 	The paper is structured as follows.
 	In Sec. \ref{sec: BKPwl}, we recall the BK teleportation protocol and we study in detail its loss dependence with a special focus on adaptive correlated losses.
 	The BK protocol in atmospheric channels is studied in Sec. \ref{sec:BKatmosphere}.
 	In Sec. \ref{sec:ModelAtmosphere} we describe the model of atmospheric free-space channels.
 	The adaptive and postselection strategies for improving teleportation through free-space channels are introduced in Secs. \ref{sec:Adaptive} and \ref{sec:Postselection}, respectively.
 	The influence of two-way atmospheric channels is studied in \ref{sec:DualAtmosphere}.
 	A summary and some conclusions are given in Sec. \ref{sec:Summary}.

\section{Braunstein-Kimble protocol with losses}
\label{sec: BKPwl}

	In this section, we will provide the framework for our analysis of the influence of atmospheric losses on CV teleportation.
	Therefore, we will briefly recall the BK protocol \cite{Braunstein98} for the teleportation of CV quantum states.
	Furthermore, we show how the teleportation fidelity---the figure of merit in such a protocol---is influenced by constant losses.
	We demonstrate that introducing additional losses can lead to an improved teleportation fidelity under certain circumstances such as strong squeezing or high losses.

\subsection{Basic protocol}

\begin{figure}[h]
		\begin{centering}
			\includegraphics[width=1\columnwidth]{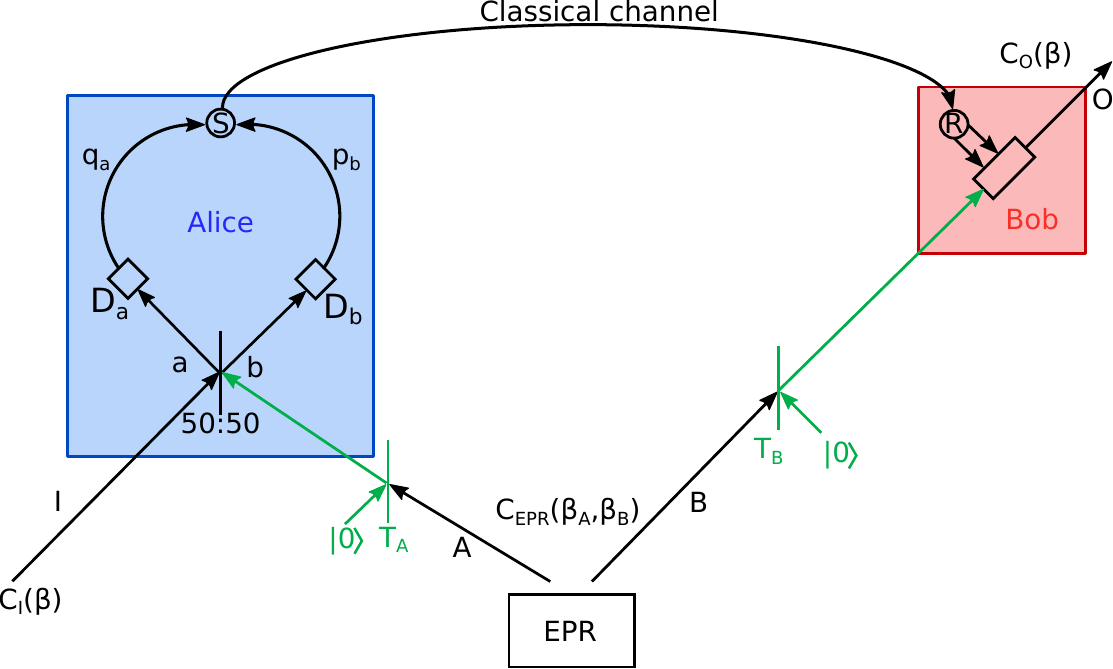}
		\par\end{centering}
		\caption{
			\label{fig: Scheme}
				Braunstein-Kimble teleportation scheme including losses.
				The aim is to teleport the input state $C_{I}\left(\beta\right)$ from Alice to Bob.
				The two parties, Alice and Bob, share the two entangled modes A and B of an two-mode squeezed vacuum state (EPR state).
				Losses in mode A(B) are characterized by the transmission coefficients $T_{\mathrm{A(B)}}$.
				Alice combines mode A with the input state at a $50:50$ beam splitter and performs a quadrature measurements of the resulting modes a and b via the detectors $\mathrm{D}_{\mathrm{a}}$ and $\mathrm{D}_{\mathrm{b}}$, respectively.
				The measurement results, $q_\mathrm{a}$ and $p_\mathrm{b}$, are sent via a classical communication channel to Bob, who modifies mode B according to this results in order to obtain the teleported output state.
		}
	\end{figure}

	In figure \ref{fig: Scheme}, the basic scheme of the BK teleportation protocol \cite{Braunstein98} is shown.
	It exists of two parties, Alice and Bob, who want to teleport an unknown state from Alice to Bob.
	The unknown input state at Alice's side is described by the characteristic function $C_\mathrm{I}\left(\beta\right)$.
	Note that the characteristic function is the Fourier transform of the Wigner function $W_\mathrm{I}(\alpha)$ and contains all the information about the quantum state.
	In the reminder of this paper, we will work with characteristic functions rather than with Wigner functions due to technical convenience; cf. also \cite{Marian06}.
	We denote the teleported output state at Bob's side as $C_\mathrm{O}\left(\beta\right)$.
	To perform the teleportation Alice and Bob need to share an entangled quantum state, which is sent along the modes $A$ and $B$, cf. figure \ref{fig: Scheme}.
	In the BK protocol, the entangled state is a two-mode squeezed vacuum or Einstein-Podolsky-Rosen (EPR) state which can be described by the characteristic function
	\begin{equation}
		C_{\mathrm{EPR}}\left(\beta_{\mathrm{A}},\beta_{\mathrm{B}}\right)= 
		\exp\left\{ 
		\begin{pmatrix}
		\beta_{\mathrm{A}}^{\ast} & \beta_{\mathrm{A}} & \beta_{\mathrm{B}}^{\ast} & \beta_{\mathrm{B}}
		\end{pmatrix}
		\boldsymbol{V}
		\begin{pmatrix}
		\beta_{\mathrm{A}}\\
		\beta_{\mathrm{A}}^{\ast}\\
		\beta_{\mathrm{B}}\\
		\beta_{\mathrm{B}}^{\ast}
		\end{pmatrix}\right\}
	\end{equation}
	with the covariance matrix
	\begin{eqnarray}
		\boldsymbol{V}=\begin{pmatrix}\cosh\left(2r\right) & 0 & 0 & -\sinh(2r)\\
		0 & \cosh\left(2r\right) & -\sinh(2r) & 0\\
		0 & -\sinh(2r) & \cosh\left(2r\right) & 0\\
		-\sinh(2r) & 0 & 0 & \cosh\left(2r\right)
		\end{pmatrix}\;.
	\end{eqnarray}
	Here, $r$ is the squeezing parameter of the two-mode squeezed vacuum state.

	For our considerations, we extend the original proposal of Braunstein and Kimble to the realistic case of losses.
	The losses in the modes A and B are characterized via the transmission coefficients $T_{\mathrm{A}}$ and $T_{\mathrm{B}}$.
	The (intensity) losses are then given by $1-T_{\mathrm{A(B)}}^2$.
	Note that the lossless case corresponds to $T_{\mathrm{A}}{=}T_{\mathrm{B}}{=}1$.
	CV teleportation protocols under constant losses have already been studied; see, e.g., \cite{Chizhov02,Fiurasek02,Faria16}.

	In the first step, the input state $C_{I}\left(\beta\right)$ and Alice's part of the entangled state, i.e. mode A, are superimposed with the help of a 50:50 beam splitter.
	The resulting state possesses correlations between the corresponding output modes a and b, and mode B; cf. figure \ref{fig: Scheme}.
	Now, Alice conducts homodyne measurements of the quadratures $q_{a} = \left(q+q_A\right)/\sqrt{2} $ and $p_{b} = \left(p-p_A\right)/\sqrt{2} $ and sends her results via a classical communication channel to Bob.
	With Alice's results Bob performs the coherent displacement of mode B as $\alpha_\mathrm{B}\mapsto\alpha_\mathrm{B}+\sqrt{2}(q_a+ip_b)$.
	The output state then reads as
	\begin{eqnarray}\label{eq:Cout}
		C_{\mathrm{O}}\left(\beta;r,T_{\mathrm{A}},T_{\mathrm{B}}\right)= & C_{\mathrm{I}}\left(\beta\right)C_{\mathrm{G}}\left(\beta^{\ast},\beta;r,T_{\mathrm{A}},T_{\mathrm{B}}\right)\;.
	\end{eqnarray}
	Here, $C_{\mathrm{G}}\left(\beta_\mathrm{A},\beta_\mathrm{B};r,\mathrm{T_{A}},T_{\mathrm{B}}\right)$ is a Gaussian state of the form
	\begin{eqnarray}
		&C_{\mathrm{G}}\left(\beta_\mathrm{A},\beta_\mathrm{B};r,\mathrm{T_{A}},T_{\mathrm{B}}\right)\\&=\exp\left\{ -\dfrac{1}{4}\begin{pmatrix}\beta_\mathrm{A}^{\ast} & \beta_\mathrm{A} & \beta_\mathrm{B}^{\ast} & \beta_\mathrm{B}\end{pmatrix}\boldsymbol{V}\left(r,T_{\mathrm{A}},T_{\mathrm{B}}\right)
		\begin{pmatrix}\beta_\mathrm{A}\\
		\beta_\mathrm{A}^{\ast}\\
		\beta_\mathrm{B}\\
		\beta_\mathrm{B}^{\ast}
	\end{pmatrix}\right\}\nonumber
	\end{eqnarray}
	with $\boldsymbol{V}\left(r,T_{\mathrm{A}},T_{\mathrm{B}}\right)$ being a $4\times4$ matrix.
	We can express $\boldsymbol{V}\left(r,T_{\mathrm{A}},T_{\mathrm{B}}\right)$ in terms of three $2\times2$ matrices $\boldsymbol{A}$,$\boldsymbol{B}$ and $\boldsymbol{C}$:
	\begin{eqnarray}
		\boldsymbol{V}\left(r,T_{\mathrm{A}},T_{\mathrm{B}}\right)=\begin{pmatrix}\boldsymbol{A} & \boldsymbol{C}\\
		\boldsymbol{C}^{\dagger} & \boldsymbol{B}
	\end{pmatrix}.
	\end{eqnarray}
	The three $2\times2$ matrices are
	\begin{eqnarray}\label{eq:A}
		\boldsymbol{A}= & \left[T_{\mathrm{A}}^{2}\cosh\left(2r\right)+\left(1-T_{\mathrm{A}}^{2}\right)\right]\begin{pmatrix}1 & 0\\
		0 & 1
		\end{pmatrix}\;,\\ \label{eq:B}
		\boldsymbol{B}= & \left[T_{\mathrm{B}}^{2}\cosh\left(2r\right)+\left(1-T_{\mathrm{B}}^{2}\right)\right]\begin{pmatrix}1 & 0\\
		0 & 1
		\end{pmatrix}\;,\\ \label{eq:C}
		\boldsymbol{C}= & -T_{\mathrm{A}}T_{\mathrm{B}}\sinh\left(2r\right)\begin{pmatrix}0 & 1\\
		1 & 0
		\end{pmatrix}\;.
	\end{eqnarray}
	Note that the product structure on the right-hand side of  \eqref{eq:Cout} corresponds to a convolution of the Wigner function of the input state with a Gaussian function.
	This additional Gaussian factor resembles noise which impairs the teleportation and is determined by the finite squeezing strength $r$ and the losses in the system characterized by $T_{\mathrm{A}}$ and $T_{\mathrm{B}}$.

	In order to characterize the teleportation quality, one usually considers the fidelity \cite{Jozsa1994, Schumacher1995} between original and teleported states.
	The fidelity quantifies the similarity of two quantum states and takes a value between $0$ and $1$ corresponding to orthogonal and identical states, respectively.
	In the case of coherent state teleportation, teleportation fidelities larger than $1/2$ can only be achieved with usage of nonclassical resources such as quantum entanglement \cite{Furusawa1998, Braunstein2000,Braunstein2001,Hammerer05}.
	Hence, the exceedance of this value can be considered as a manifestation of quantum advantages for the task of quantum teleportation.
	The teleportation fidelity is the figure of merit for teleportation protocols.
	In the case of pure input state, $\hat{\rho}_\mathrm{I}$, the fidelity with the output state $\hat{\rho}_\mathrm{O}$ is defined as $F=\mathrm{Tr}\left(\hat{\rho}_\mathrm{I}\hat{\rho}_\mathrm{O}\right)$.
	In terms of the corresponding characteristic functions this relation is given by
	\begin{eqnarray}\label{eq:Fidelity}
		F\left(r,T_{\mathrm{A}},T_{\mathrm{B}}\right)=\dfrac{1}{\pi}\int d^{2}\beta\,C_{\mathrm{I}}\left(\beta\right)\,C_{\mathrm{O}}\left(-\beta;r,T_{\mathrm{A}},T_{\mathrm{B}}\right).
	\end{eqnarray}
	For the considered scenario, it depends on the squeezing strength $r$ and the two transmission coefficients $T_{A(B)}$.

	As a particular case, the fidelity of coherent-state teleportation can be directly expressed in terms of the blocks of the covariance matrix \cite{Chizhov02,Fiurasek02} given in  \eqref{eq:A}-\eqref{eq:C},
	\begin{eqnarray}
		F=\frac{2}{\sqrt{\det \boldsymbol{E}}},
	\end{eqnarray}
	with
	\begin{eqnarray}
		\boldsymbol{E}=2 \boldsymbol{I}+\boldsymbol{RAR}+\boldsymbol{C}^\dagger \boldsymbol{R}+\boldsymbol{RC}+\boldsymbol{B},
	\end{eqnarray}
	were $\boldsymbol{I}$ is the two-dimensional identity matrix, and
	\begin{eqnarray}
		\boldsymbol{R}=
		\begin{pmatrix}
		1 & 0\\
		0 & -1
		\end{pmatrix}.
	\end{eqnarray}
	All the relevant information for the teleportation process about the entangled state and the channel losses are included in the covariance-matrix elements.
	Eventually, this yields the analytical expression for the teleportation fidelity including losses,
	{\fontsize{9}{8}\selectfont
	\begin{eqnarray}\label{eq:Fidelity2}
		F&\left(r,T_{\mathrm{B}},T_{\mathrm{A}}\right)=\nonumber\\
		&\dfrac{2}{4+\left(T_{\mathrm{A}}^{2}+T_{\mathrm{B}}^{2}\right)\left(\cosh\left(2r\right)-1\right)-2T_{\mathrm{A}}T_{\mathrm{B}}\sinh\left(2r\right)} \;.
	\end{eqnarray}
	}

	As long as the fidelity exceeds the classical limit of teleportation, $F_{\mathrm{cl}}=0.5$, the protocol shows a quantum advantage in the considered scenario \cite{Hammerer05}.
	As we are interested in the quantum advantage of the considered teleportation scenarios, we will plot the fidelities in this paper only from $0.4$ upwards.
	If the fidelity is below $0.5$ there is no quantum advantage in the considered strategy.
	Note that the fidelity is a function of the squeezing parameter $r$ and the two transmission coefficients $T_{\mathrm{A}}$ and $T_{\mathrm{B}}$.
	In the following we will analyze the dependence of the teleportation fidelity on (atmospheric) losses and develop strategies for minimizing the unwanted effects of these losses on the teleportation result.

\subsection{Loss dependence and adaptive protocol}
	Let us now discuss the influence of constant losses on the teleportation fidelity in  \eqref{eq:Fidelity2}.
	We will analyze two cases.
	In the first case, we only consider losses in mode B and set the transmission coefficient $T_{\mathrm{A}}$ to unity.
	We will call this the direct scheme, in which Alice may control mode A in a lossless loop.
	In the second case, the transmission coefficient $T_{\mathrm{A}}$ will be set to the value of the transmission coefficient $T_{\mathrm{B}}$.
	This case will be called the adaptive scheme as the loss in mode A is adapted to be the same loss as in mode B.

	\begin{figure}[ht]
		\begin{centering}
			\includegraphics[width=1\columnwidth]{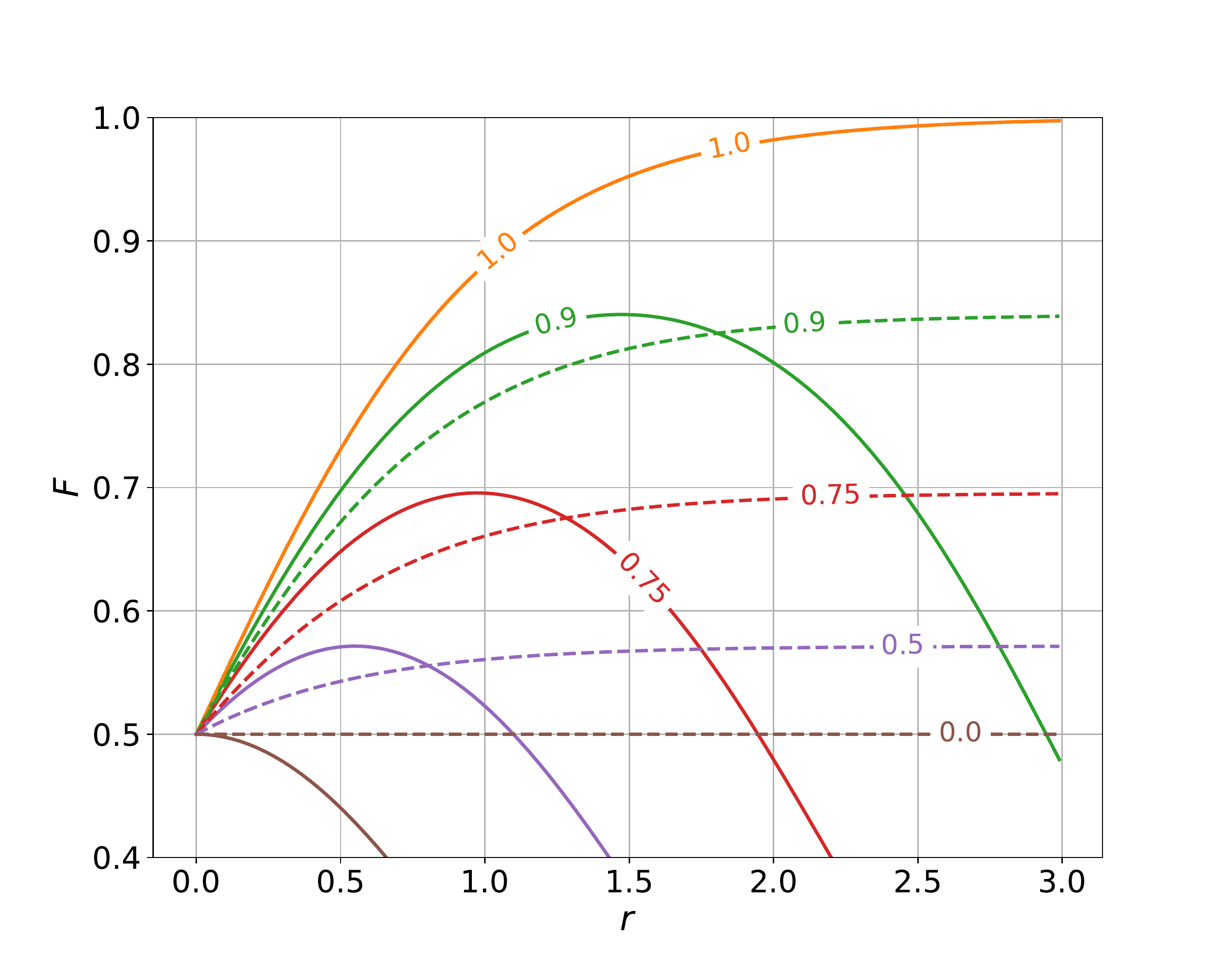}
		\par\end{centering}
		\caption{
			\label{fig: F}
			Fidelity in dependence on the squeezing parameter.
			For the solid lines, $T_{\mathrm{B}}$ is varied and $T_{\mathrm{A}}$ is set to $1.0$.
			For the dashed lines, $T_{\mathrm{B}}$ is varied and $T_{\mathrm{A}}$ is set to $T_{\mathrm{B}}$.
			The curves are labeled with the corresponding value of the transmission coefficient $T_{\mathrm{B}}$.
		}
	\end{figure}

	In figure \ref{fig: F}, the teleportation fidelity is plotted in dependence on the squeezing parameter $r$, for different losses for the direct and adaptive schemes.
	For the direct scheme (solid lines), the fidelity grows from $0.5$ to a maximum and than eventually drops to zero for increasing values of the squeezing parameter.
	Only in the lossless case $(T_{\mathrm{A}}{=}T_{\mathrm{B}}{=}1)$, the fidelity approaches unity with increasing squeezing parameter.
	Hence, we observe that in the direct scheme with losses there exists an optimal and finite squeezing value for which the best teleportation fidelity is obtained.
	The value of this best teleportation fidelity and its corresponding squeezing value depend on the transmission coefficient $T_{\mathrm{B}}$.
	A higher transmission coefficient $T_{\mathrm{B}}$ leads to a higher possible fidelity but requires stronger squeezing.
	The optimal squeezing parameter which yields the maximal fidelity in the lossy case is given by
	\begin{eqnarray}\label{eq:ropt}
		r_{\mathrm{opt}}\left( T_{\mathrm{A}} , T_{\mathrm{B}} \right)=\dfrac{1}{2} \mathrm{arctanh}\left({\dfrac{2 T_{\mathrm{A}} T_{\mathrm{B}}}{T_{\mathrm{A}}^2 + T_{\mathrm{B}}^2}}\right)\;,
	\end{eqnarray}
	which reduces to
	\begin{eqnarray}
		r_{\mathrm{opt}}\left( T_{\mathrm{B}} \right)=\dfrac{1}{2} \mathrm{arctanh}\left({\dfrac{2 T_{\mathrm{B}}}{1 + T_{\mathrm{B}}^2}}\right)\;,
	\end{eqnarray}
	in the case of losses in mode B only.
	From \eqref{eq:ropt} we clearly see that in the uncorrelated-loss case $(T_{\mathrm{A}}{\neq}T_{\mathrm{B}})$ there exists an optimal finite squeezing value for which the highest teleportation fidelity is reached.
	Increasing the squeezing parameter beyond $r_{\mathrm{opt}}$ leads to a reduction of the teleportation fidelity; cf. figure \ref{fig: F}.
	Similar behaviors can also be observed in the case of fluctuating losses (see figures \ref{fig: ada_and_atm} and \ref{fig: dual atmo}) as we will show below.
	
	This can be explained by the fact that higher squeezing leads to a higher mean photon number in both modes.
	Higher photon-number contributions are, however, more sensitive towards losses as the $n$-th photon-number contribution scales with the $n$-th power of the corresponding transmission coefficient.
	Therefore, increasing the squeezing parameter leads to an overall stronger influence of the losses which manifests itself in a more pronounced asymmetry of the state.
	The symmetry in the state is, however, essential for the quantum advantage in the teleportation.
	This explains why there is a finite optimal squeezing value in the direct scheme.
	More formally, this effect can be understood by the factor $C_{\mathrm{G}}\left(\beta_\mathrm{A},\beta_\mathrm{B};r,\mathrm{T_{A}},T_{\mathrm{B}}\right)$ in the input-output relation~(\ref{eq:Cout}) which is a characteristic function of a Gaussian distribution.
	The performance of the teleportation increases when the variance of this Gaussian distribution is decreasing.
	Note that in the ideal EPR case the distribution approaches a delta distribution with zero variance for infinite $r$ \cite{Braunstein98}.
	For asymmetric losses $(T_{\mathrm{A}}{\neq}T_{\mathrm{B}})$, the variance attains its minimum at the finite value $r_{\mathrm{opt}}$.
	
	Next, we consider the adaptive case in which the losses in mode A are adapted to be the same as in mode B, i.e., $T_{\mathrm{A}}{=}T_{\mathrm{B}}{=}T$.
	Such an adaptive protocol can always be realized by measuring the transmission coefficient in one channel and then artificially attenuate the other channel to the measured level \cite{BSSV16}.
	A similar adaptive scenarios for quantum teleportation has been proposed in \cite{Ban06}, however, from a different perspective.
	For different amounts of losses, the corresponding values of the teleportation fidelities are plotted in figure \ref{fig: F} (dashed lines). 
	In this case, the fidelity in \eqref{eq:Fidelity2} reduces to 
	\begin{eqnarray}
		F(r,T) = \dfrac{1}{2-T^2(1-\exp(-2r))}.
	\end{eqnarray}
	In contrast to the direct case, in the adaptive scheme the fidelity increases monotonically with increasing squeezing parameter $r$ to the upper limit
	\begin{eqnarray}
		F_{\mathrm{opt}} = \dfrac{1}{2-T^2}.
	\end{eqnarray}
	For an increasing transmission coefficient $T$ this upper limit also increases, but it is never higher than the maximum in the direct scheme.
	However, for a fixed squeezing value the adaptive protocol can yield higher teleportation fidelities in comparison to the direct scheme especially for higher losses.
	The crossing point of the solid and dashed lines in figure \ref{fig: F}, i.e., the squeezing strength from which on the adaptive protocol performs better than the direct teleportation, is given by
	\begin{eqnarray*}
		r\left( T_\mathrm{B} \right)=\mathrm{arctanh}\left(\frac{2 T_\mathrm{B}}{1+T_\mathrm{B}}\right)\;.
	\end{eqnarray*}
	Furthermore, the fidelity in the adaptive scenario never drops below the classical limit of $0.5$, which does not hold for the direct scheme.
	Hence, we could show that establishing correlations in the losses can improve the performance of the teleportation even though this implies to introduce additional losses.
	It is important to stress that the correlations between the modes, including correlations in the losses, can be more important than the overall value of the transmittance of the channel.

\section{Braunstein-Kimble protocol with atmospheric channels}
\label{sec:BKatmosphere}

	In this section, we consider the action of atmospheric losses on the teleportation protocol.
	Therefore, we recall a model for the description of atmospheric losses and apply this model to the BK teleportation protocol.
	We extend the direct and adaptive schemes to the regime of atmospheric losses and analyze the corresponding fidelities.
	Furthermore, we show how postselection strategies can lead to an improved teleportation fidelity.
	Finally, we study the case of two-way atmospheric channels in which both entangled modes suffer from uncorrelated atmospheric losses.

\subsection{Modeling atmospheric losses}
\label{sec:ModelAtmosphere}

	In quantum optics, losses can be described by a virtual beam splitter that superimposes the lossless state with vacuum noise.
	The transmission coefficient of the beam splitter depends on the losses.
	Fluctuating atmospheric channels can be modeled by such a beam splitter for which the transmission coefficient is a random variable \cite{Semenov2009}.
	The properties of the channel is then characterized by the probability distribution of the transmittance (PDT) \cite{beamwandering,VSV2016,Vasylyev17,Vasylyev18}.

	For a start, we consider atmospheric losses only in mode B.
	The output state of the quantum channel is obtained by averaging the fidelity in  \eqref{eq:Fidelity2} over the atmospheric PDT $\mathcal{P}\left(T_{\mathrm{B}}\right)$.
	This means that the fidelity of the teleportation should also be averaged as
	\begin{eqnarray}
		\bar{F}(r)=\int_{0}^{1}dT_{\mathrm{B}}F\left(r,T_{\mathrm{A}},T_{\mathrm{B}}\right)\mathcal{P}\left(T_{\mathrm{B}}\right)\;.
	\end{eqnarray}
	Note that $\bar{F}(r)$ directly depends on the PDT $\mathcal{P}\left(T_{\mathrm{B}}\right)$ and, hence, on the properties of the atmospheric channel.
	This treatment hold for any turbulent free-space channel.

	In the following, we exemplarily work with the PDT of the elliptic beam model \cite{VSV2016}.
	This model takes into account the deflection and deformation of a Gaussian beam caused by turbulence in atmospheric channels and shows good agreement with experimental free-space channels.
% 	The atmospheric PDT than depends on the overlap of the beam with the aperture at the destination.
	In particular, we consider the example of a $1.6$km long free-space link which has been implemented in the city of Erlangen in Germany \cite{Usenko}.
	The detailed description of this channel can be found in \cite{VSV2016} and the implementation of the model for calculating the teleportation fidelity is explained in the \ref{sec:Appendix}.

\begin{figure}[ht]
		\begin{centering}
			\includegraphics[width=1\columnwidth]{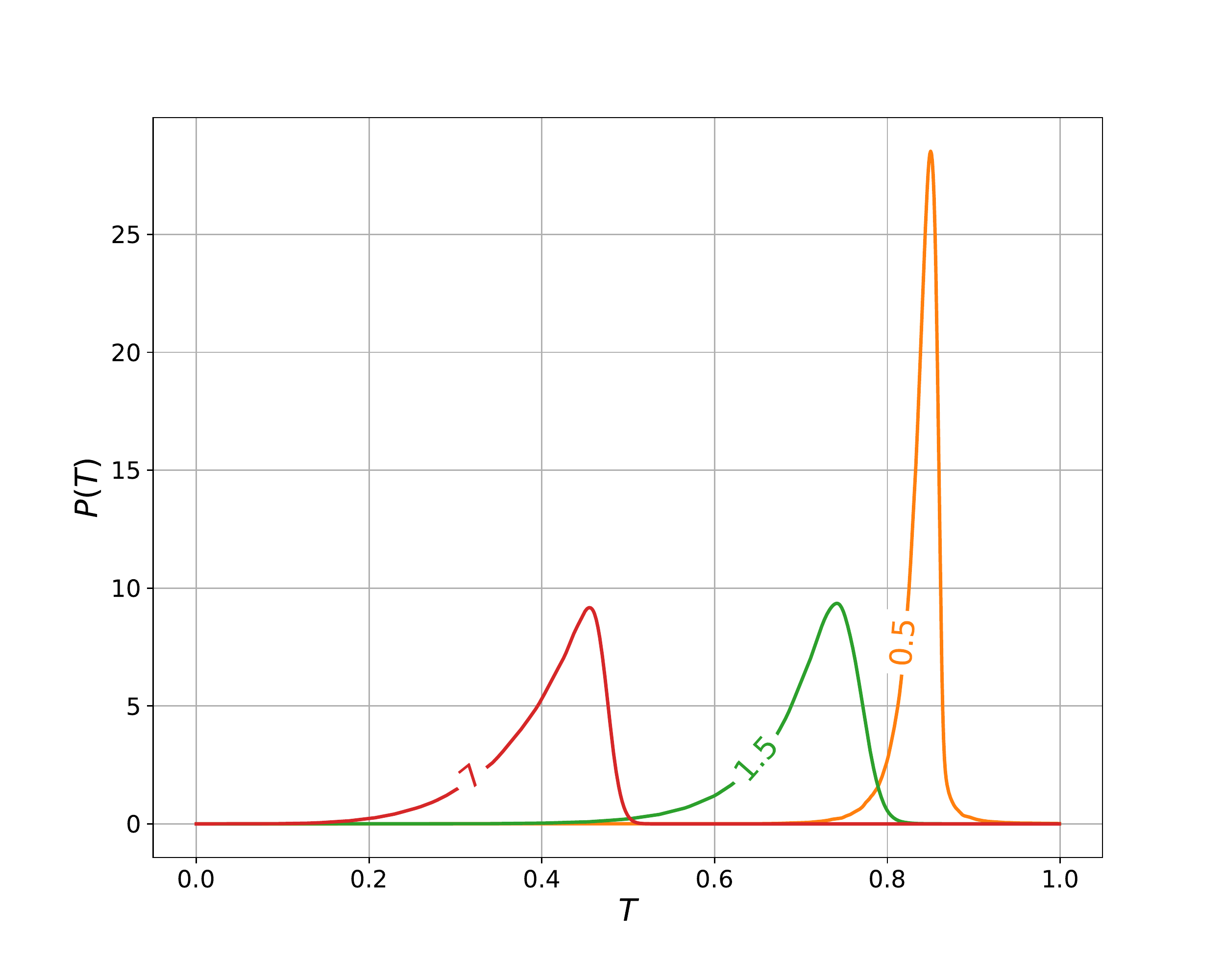}
		\par\end{centering}
		\caption{
			\label{fig: atmo_prob}
			Probability distribution of the atmospheric transmission coefficent given by the elliptic beam model \cite{VSV2016}, for different atmospheric index-of-refraction structure constants $C_\mathrm{n}^2$. The labels give the atmospheric index-of-refraction structure constants in units of $10^{-14}\;\mathrm{m^{-\frac{2}{3}}}$. Further channel parameters are given in \ref{sec:Appendix}.
		}
	\end{figure}

	In figure~\ref{fig: atmo_prob}, PDTs are shown for three different turbulence strengths, characterized by the index-of-refraction structure constants $C_\mathrm{n}^2$ \cite{Fante,Tatarskii,Andrews}.
	The corresponding distributions have average transmission coefficients $\langle T\rangle$ of $0.40$, $0.70$ and $0.84$.
	Both distributions with the lower mean transmission coefficients, corresponding to stronger turbulence, have nearly the same but displaced PDTs with a standard deviation of $\sqrt{\langle T^2\rangle-\langle T\rangle^2}=0.062$.
	The distribution with the highest expectation value has a standard deviation of $0.024$ and is therefore much thinner.
	It is important to stress that the following considerations apply to any PDT model.

\subsection{The direct and adaptive schemes}
\label{sec:Adaptive}

\begin{figure}[ht]
		\begin{centering}
			\includegraphics[width=1\columnwidth]{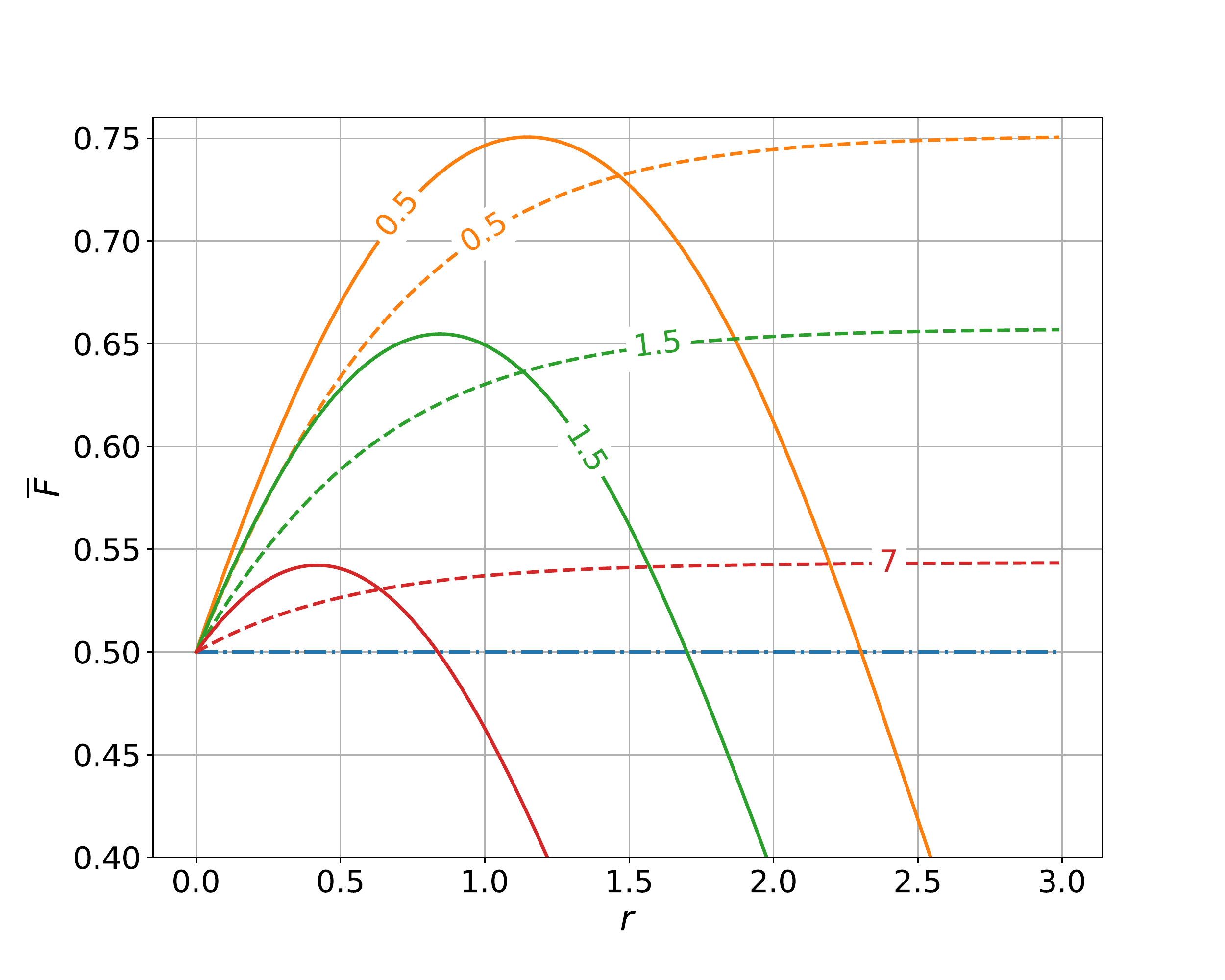}
		\par\end{centering}
		\caption{
			\label{fig: ada_and_atm}
			Average fidelity $\bar{F}$ in dependence on the squeezing parameter $r$ for the probability distributions in figure \ref{fig: atmo_prob}.
			The solid lines result from the direct scheme and the dashed lines result from the adaptive scheme.
			The dashed-dotted line is the classical limit.
		}
	\end{figure}

	As a first step, we consider the influence of fluctuating losses in atmospheric channels on the direct and adaptive teleportation schemes.
	We study the case in which mode B passes through a turbulent free-space link while mode A does not suffer from any losses, i.e. $T_{\mathrm{A}}{=}1$.
	For the three different turbulent strengths [cf. figure \ref{fig: atmo_prob}], the fidelity of the direct teleportation protocol is plotted in figure \ref{fig: ada_and_atm} (solid lines) in dependence on the squeezing parameter.
	For this direct case, we observe a similar behavior as discussed for the case of constant losses, cf. also figure \ref{fig: F}.
	For the adaptive channel protocol, the channel transmittance in mode B has to be  monitored and the measured transmittance has to be constantly adapted in mode A.
	Such an atmospheric adaptive channel protocol has been introduced in \cite{BSSV16}.
	The measurement of the channel transmittance can be directly implemented in the procedure of balanced homodyne detection \cite{Elser2009, Heim2010, Semenov2012b} or can be measured independently with an intense reference light pulse.

	The average fidelity in dependence on the squeezing parameter $r$ for the direct and adaptive schemes with different turbulence strengths (different $C_\mathrm{n}^2$) is shown in figure \ref{fig: ada_and_atm}.
	A lower atmospheric index-of-refraction structure constant, $C_\mathrm{n}^2$, results in a higher average fidelity for both schemes.
	Similar to the case of constant losses, the adaptive scheme does not improve the maximum average fidelity.
	Furthermore, as seen in figure \ref{fig: F}, the fidelity in the adaptive scheme will never drop below the classical limit.
	Therefore, the adaptive scheme is preferable in the case when the quantum channel is realized with the relatively strong squeezing.

\subsection{Postselection scheme}
\label{sec:Postselection}

	The teleportation fidelity through free-space links can be further improved by postselecting the events with high transmission coefficients.
	Such a procedure has been theoretically analyzed for improving the transmission of quadrature squeezing in \cite{beamwandering} and experimentally realized in \cite{Peuntinger}.
	Here we will study its applicability for the BK-CV teleportation protocol.

	Let us assume that we postselect only the events with the transmission coefficients $T_\mathrm{B}\geq T_\mathrm{min}$, where $T_\mathrm{min}$ is a certain postselection threshold.
	In this case the PDT is modified to the form
		\begin{eqnarray}
		\mathcal{P}_\mathrm{ps}\left(T_{\mathrm{B}};T_{\mathrm{min}}\right)=\frac{1}{\mathcal{E}(T_\mathrm{min})}\left\{\begin{array}{lr}
		\mathcal{P}\left(T_{\mathrm{B}}\right) & T_{\mathrm{B}}\geq T_{\mathrm{min}}\\
		0 & T_{\mathrm{B}}<T_{\mathrm{min}}
		\end{array}
		\right.,
		\end{eqnarray}
	where $\mathcal{E}(T)=\int_{T}^1d T^\prime \mathcal{P}(T^\prime)$ is the PDT exceedance  (complementary cumulative distribution function), which describes the total efficiency of the postselection for a given $T_\mathrm{min}$.
	A higher value of $T_\mathrm{min}$ implies that more data is discarded.
	We can apply the postselection to the direct and adaptive scheme by replacing the atmospheric probability distribution to improve the teleportation fidelity.

\begin{figure}[ht]
	\begin{centering}
		\includegraphics[width=1\columnwidth]{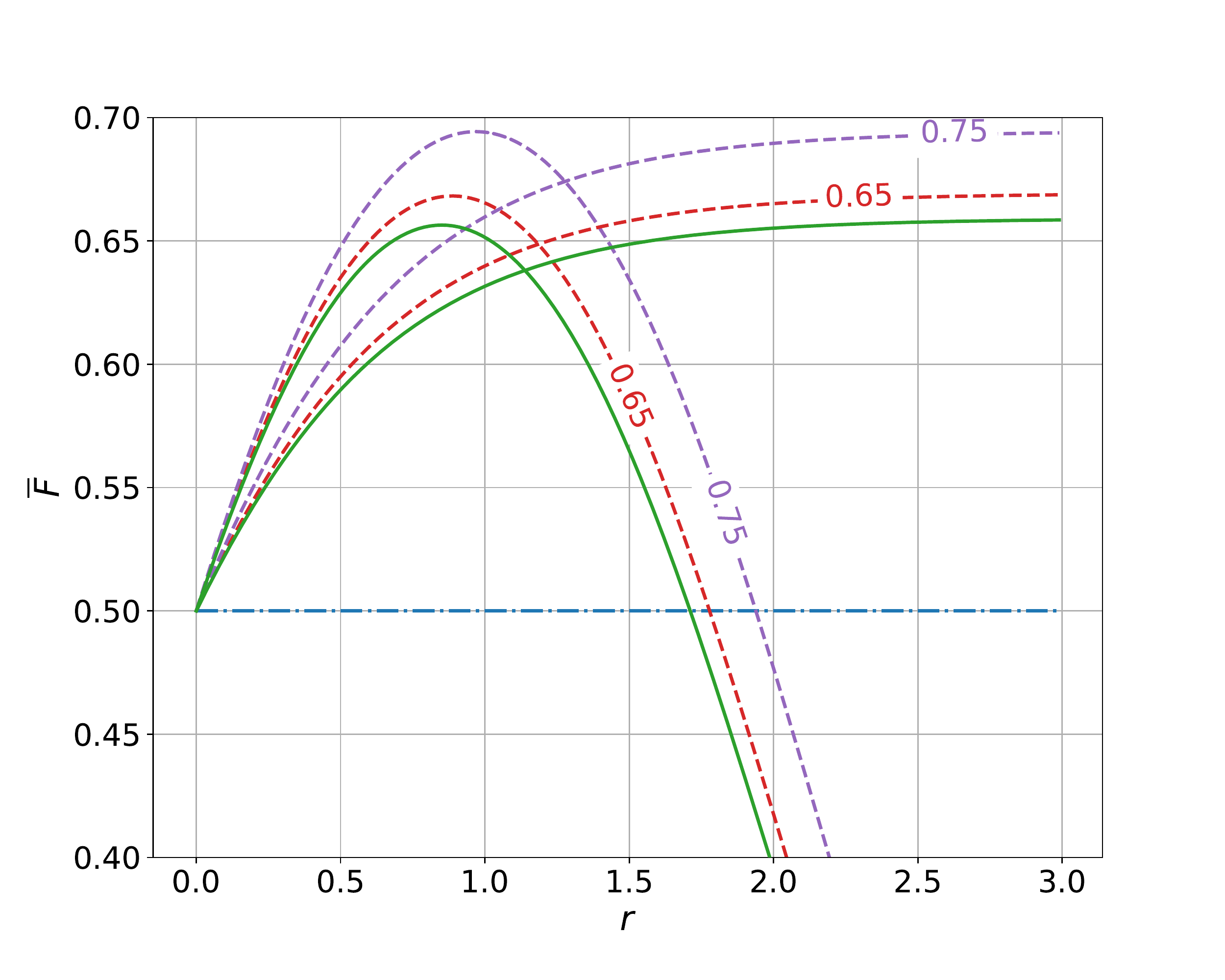}
	\par\end{centering}
	\caption{
		\label{fig: sawp}
		The average fidelity $\bar{F}$ in dependence on the squeezing parameter $r$ for a $C_\mathrm{n}^2$ of $1.5\cdot 10^{-14}\;\mathrm{m^{-\frac{2}{3}}}$ is shown.
		The monotone increasing curves result from the adaptive scheme.
		Dashed curves are with added postselection.
		The labeled numbers represent the chosen postselect threshold $T_\mathrm{min}$.
		The dashed-dotted line shows the classical limit.
	}
\end{figure}

	In figure \ref{fig: sawp}, we show the average fidelity in dependence on the squeezing parameter $r$ for the direct and adaptive schemes with postselection.
	For the sake of simplicity, we limit our plots to the case of $C_\mathrm{n}^2=1.5\cdot 10^{-14}\;\mathrm{m^{-\frac{2}{3}}}$.
	For different $C_\mathrm{n}^2$, the main features of the plot will not change.

	\begin{figure}[ht]
		\begin{centering}
			\includegraphics[width=1\columnwidth]{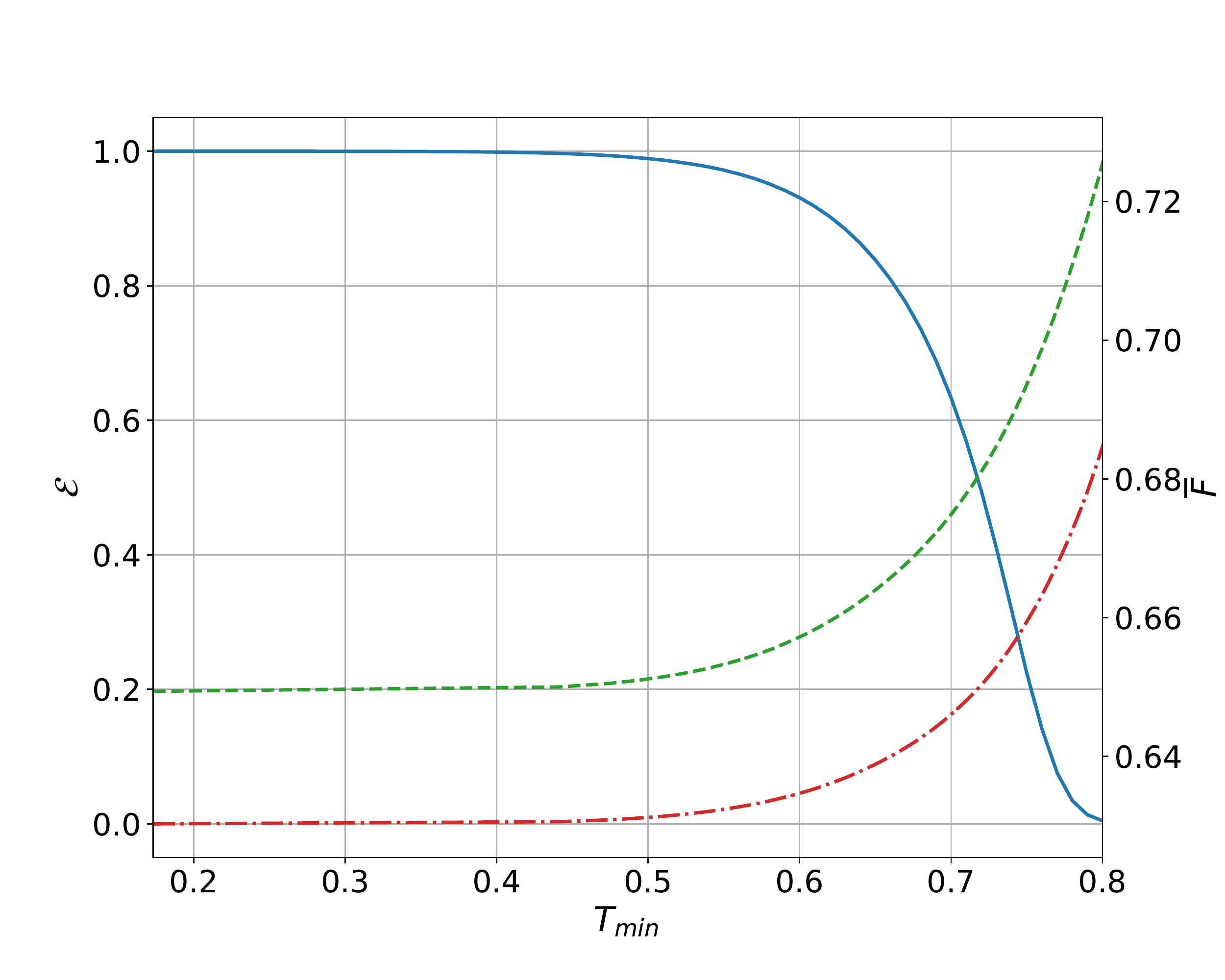}
		\par\end{centering}
		\caption{
			\label{fig: postselection_eff}
			The solid (blue) line shows the postselection efficiency, i.e., the PDT exceedance $\mathcal{E}$, (left axis) in dependence on the postselect threshold $T_{\mathrm{min}}$.
			The dashed (green) line and the dashed-dotted (red) line show the fidelity (right axis) in dependence on $T_{\mathrm{min}}$ for the adaptive and direct scheme for a squeezing parameter $r=1$, respectively.
		}
	\end{figure}

	We observe that the postselection improves the average fidelity and that it increases with the postselection threshold $T_\mathrm{min}$.
	This improvement, however, comes with the disadvantage that the postselection procedure implies that we have to discard part of the data.
	In figure \ref{fig: postselection_eff}, the postselection efficiency $\mathcal{E}(T_{\mathrm{min}})$ is shown together with the average fidelity for the postselected direct and adaptive schemes in dependence on the postselection threshold $T_{\mathrm{min}}$.
	We observe that the teleportation fidelity can be improved by means of postselection if one is willing to reduce the teleportation efficiency due to the reduction of the amount of teleported data.
	This is an important finding which allows to improve the teleportation performance in free-space channels.
	Note that by postselection it is even possible to increase the fidelity from classically achievable values ($\bar{F}\leq0.5$) to values which show a quantum advantage ($\bar{F}>0.5$), as can be seen in figure \ref{fig: sawp}.
	Importantly, the possibility of implementing such a postselection scheme is an unique feature of fluctuating-loss channels and cannot be implemented in constant-loss scenarios, such as losses in optical fibers.
	Hence, free-space channels with postselection can allow for quantum teleportation ($\bar{F}>0.5$) in cases when alternative implementations with constant losses fail.

\subsection{Two-way atmospheric channels}
\label{sec:DualAtmosphere}

	We can extend the consideration of teleportation through atmospheric channels to the case where both modes, A and B, suffer from (uncorrelated) atmospheric losses.
	In this case, the transmission coefficient $T_{\mathrm{A}}$ represents the atmospheric transmission coefficient of mode A which also fluctuates in a random fashion.
	In the following, we consider the case in which both modes propagate through 1.6km-long atmospheric channels \cite{Usenko, VSV2016}.
	Similar to the previous consideration, we can apply an adaptive scheme to the protocol.
	Here the adaptive scheme means, that we adjust the higher atmospheric transmission coefficient to the lower atmospheric transmission coefficient.
	We can furthermore apply postselection.
	In this case we do not teleport if one of the atmospheric transmissions is lower than a given limit.

	The joint PDT, $\mathcal P'$, of our adaptive scheme can be obtained straightforwardly from the initial distribution $\mathcal P$  by mapping the random variables $T_a,T_b\mapsto\min\{T_a,T_b\}$ \cite{BSSV16} within the technique of order statistics \cite{David}, 
	\begin{eqnarray}\label{eq:adaptivePDTC}
		\mathcal P'(T_a,T_b){=}\\ \nonumber
		\delta(T_a{-}T_b)\!\!\left[
			\int\limits_{T_a}^1\!\!dT'_a\mathcal P(T'_a,T_b)
			{+}\!\!
			\int\limits_{T_b}^1\!\!dT'_b\mathcal P(T_a,T'_b)
		\right]\!\!.
	\end{eqnarray}

	\begin{figure}
		\begin{centering}
			\includegraphics[width=1\columnwidth]{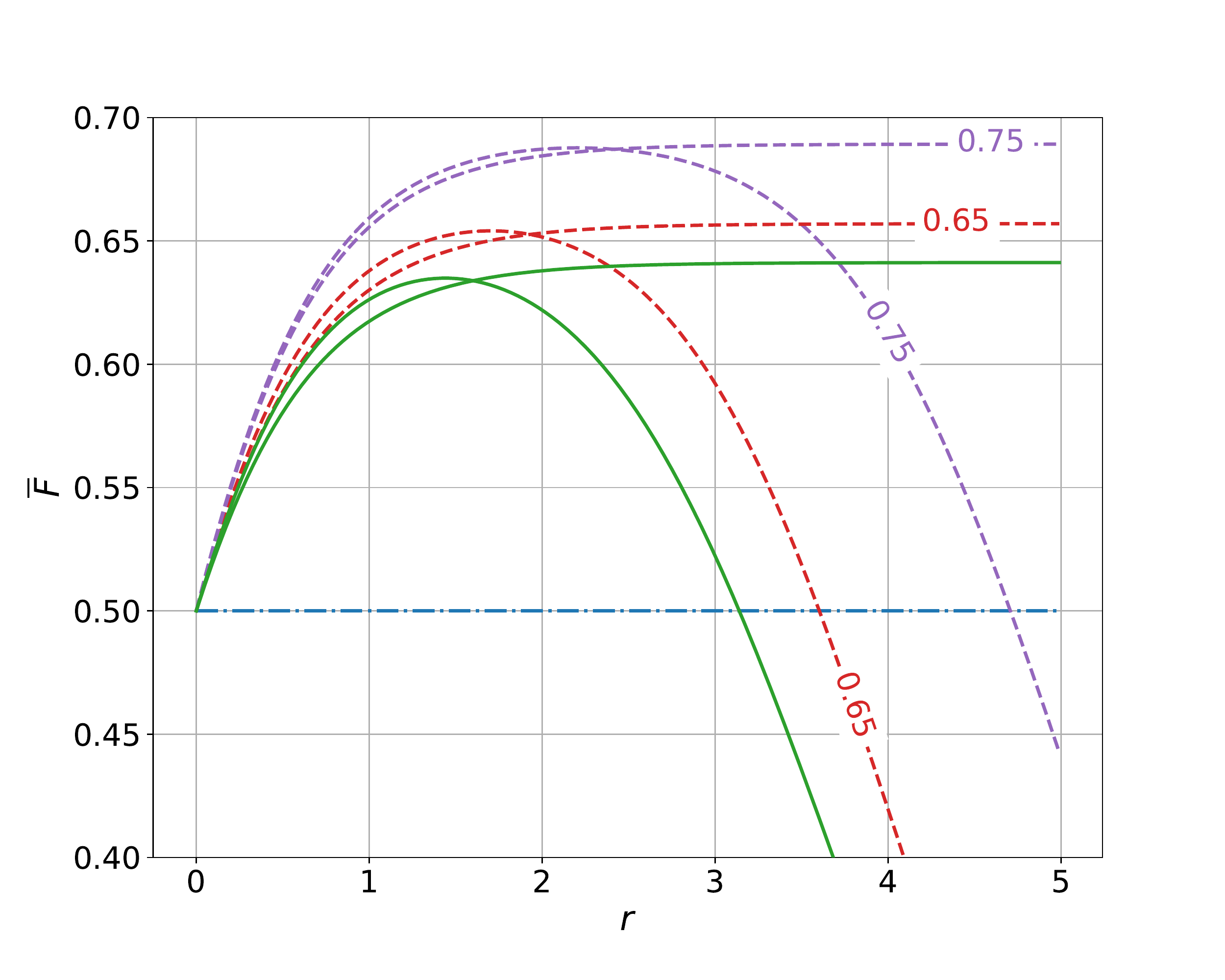}
		\par\end{centering}
		\caption{
			\label{fig: dual atmo}
			Average fidelity $\bar{F}$ in dependence on the squeezing parameter $r$ for the case of atmospheric losses in both modes A and B.
			The monotonously increasing curves result from the adaptive scheme, the others are obtained without the adaptive protocol.
			Dashed curves are with added postselection.
			The labeled numbers indicate the corresponding postselection threshold.
			The dashed-dotted line is the classical limit.
		}
	\end{figure}

	In figure \ref{fig: dual atmo}, we show the average fidelity in dependence on the squeezing parameter $r$ for the direct and adaptive schemes, with and without postselection, for a turbulence strength characterized by $C_\mathrm{n}^2$ of $1.5\cdot 10^{-14}\;\mathrm{m^{-\frac{2}{3}}}$.
	Different from the case of atmospheric noise in one of the entangled modes, the adaptive scheme can improve the fidelity above its maximal value for the uncorrelated case.
	Furthermore, we also observe that postselection improves the average fidelity.
	But the postselection efficiency is lower compared to the case of a single atmospheric channel.

	\begin{figure}[ht]
		\begin{centering}
			\includegraphics[width=1\columnwidth]{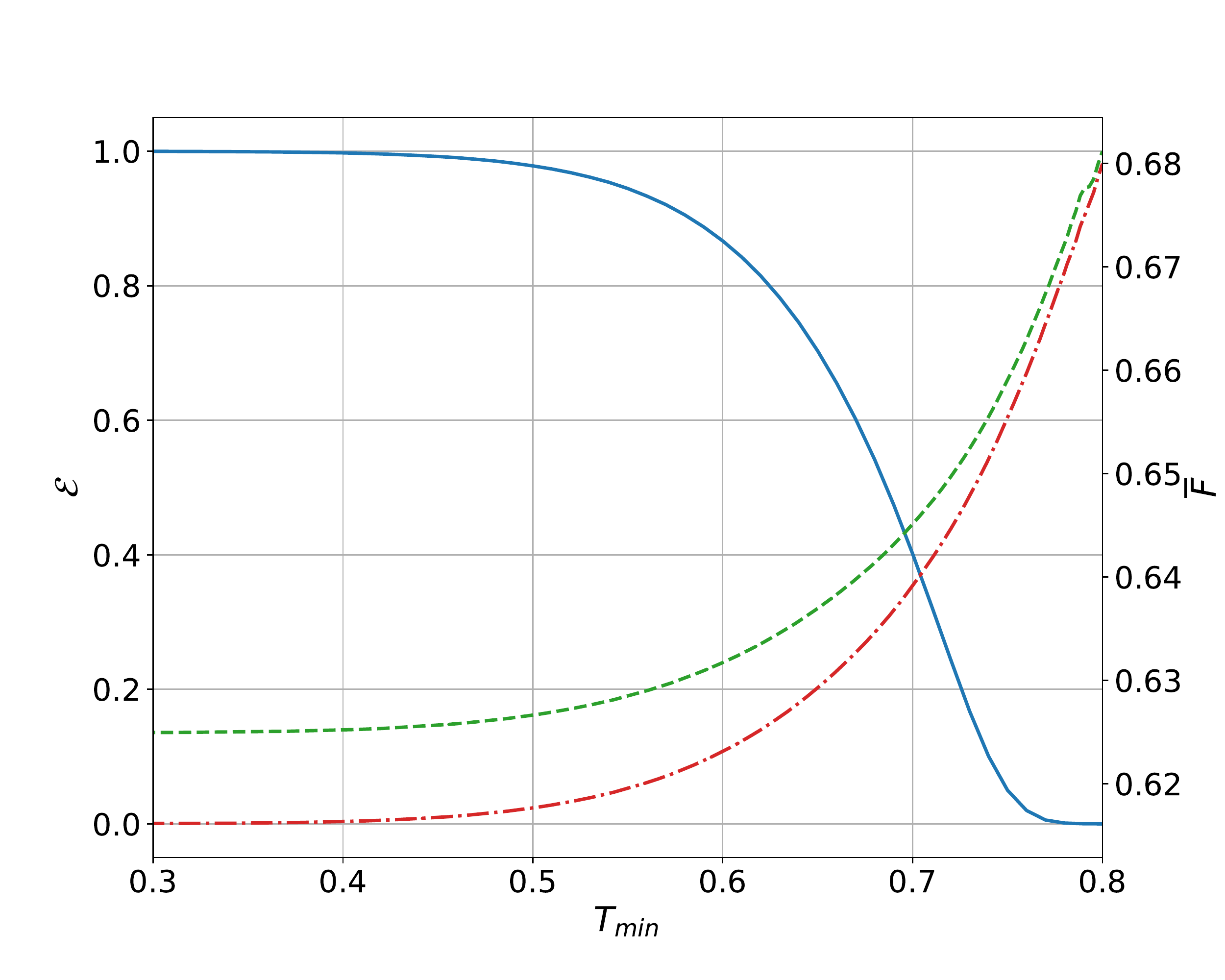}
		\par\end{centering}
		\caption{
			\label{fig: postselection_eff_dual}
			The solid (blue) line shows the postselection efficiency (left axis) in dependence on the lower limit $T_{\mathrm{min}}$ for atmospheric channels A and B.
			The dashed (green) line and the dashed-dotted (red) line show the fidelity (right axis) in dependence on the lower limit $T_{\mathrm{min}}$ for the adaptive and direct scheme, respectively, for a squeezing parameter of $r=1$.
		}
	\end{figure}
	
	As in the case of a single atmospheric channel, we show in figure \ref{fig: postselection_eff_dual} the postselection efficiency and the average fidelity as functions of the lower limit of the transmission threshold, $T_{\mathrm{min}}$, for the direct and adaptive postselection scheme.
	As before, an increment of the average fidelity results in a decrease of the postseletion efficiency.
	Also the postselection efficiency for atmosphere in both channels is lower than for atmosphere only in one channel.
	Like in the case of atmosphere only in mode B, postselection can increase the fidelity from classical achievable values $\left( \bar{F} \leq 0.5 \right)$ to values which show quantum advantages $\left( \bar{F} > 0.5 \right)$.
	We conclude that also for two-way atmospheric channels it is possible to increase the teleportation fidelity by means of adaptive loss correlation and postselection.

\section{Summary and outlook}
\label{sec:Summary}

	We analyzed the BK-CV teleportation protocol under the influence of constant and fluctuating losses.
	In particular, losses occurring in the two-mode squeezed vacuum state, used as the resource for the teleportation, were considered.
	We started our fully analytical treatment with the consideration of the influence of constant losses on such a teleportation protocol.
	For this scenario, we could show that introducing additional losses in the system can lead to an improvement in the teleportation fidelity under certain circumstances.
	This improvement stems from the introduced correlations in the losses which can outstrip the negative effect of the additional loss.
	Analytical expressions for the teleportation fidelity depending on the squeezing parameter and the loss parameters were derived together with the conditions under which the additional adaptive losses lead to an improved performance.

	After the consideration of the constant-loss case, we extended the treatment to fluctuating atmospheric losses.
	Therefore, we first recalled the theoretical description of such channels and considered three different loss distributions covering both weak and strong turbulence conditions.
	We could show that the adaptive loss-correlation technique can also be applied in the case of fluctuating losses in atmospheric free-space channels.
	Furthermore, we demonstrated that post-selection procedures can further enhance the teleportation fidelity under such conditions.
	In this context, we also analyzed the relation between the increase of the fidelity and the amount of data which has to be discarded.
	Finally, we studied the case in which both entangled modes suffer form uncorrelated fluctuating losses.
	For this scenario, we also demonstrated that the proposed techniques of adaptive loss correlations and post-selection can be beneficial for quantum state teleportation.

	We believe that our proposed strategies for the improvement of CV quantum-state teleportation through fluctuating-loss channels will help to implement such teleportation schemes under realistic conditions, which eventually will lead to practical applications.
	All proposed techniques can be directly implemented in common teleportation experiments.
	It would be interesting to extend the present consideration to other teleportation protocols and analyze which strategies are the most practical and robust ones in the presence of atmospheric losses.
	For teleportation through atmospheric channels, protocols relying on non-Gaussian entanglement might be beneficial as such state can be more robust towards fluctuating losses. 
	Furthermore, an extension to hybrid discrete-continuous variable systems might be promising, as advantages of both systems could be explored.

\ack 
	The authors are grateful to D. Vasylyev for enlightening discussions.
	This work has been supported by Deutsche Forschungsgemeinschaft through Grant No. VO 501/22-2.
	AAS also acknowledges support from the Department of Physics and Astronomy of the NAS of Ukraine through the project PK 0118U003535.
	MB acknowledges financial support by the Leopoldina Fellowship Programme of the German National Academy of Science (LPDS 2019-01).

\appendix
\section{Atmospheric channel parameters}
\label{sec:Appendix}

	In this Appendix, we briefly discuss how to apply the method of the elliptic-beam approximation \cite{VSV2016} for calculation of the mean fidelity $\bar{F}=\int_0^1d T \mathcal{P}(T)F(T)$.
	Further details on the model can be found in \cite{VSV2016}.
	For this purpose we generate $N$ independent Gaussian random vectors $\mathbf{v}_i=\big(x_{0;i}\,y_{0;i}\,\Theta_{1;i}\,\Theta_{2;i}\big)^\mathrm{T}$ and random uniformly-distributed angles $\chi_i\in[0,\pi/2]$, for $i=1\ldots N$.
	Here $x_{0;i}$ and $y_{0;i}$ are random coordinates of the beam centroid and $\Theta_{1/2;i}$ are related to the semi-axes, $W_i$, of random ellipses, which model the beam profile after transferring through the atmosphere such that $W_{1/2;i}^2=W_0^2\exp\left[-\Theta_{1/2;i}\right]$ with $W_0$ being the beam-spot radius at the transmitter.
	The non-zero elements of the covariance matrix and the means are given by
	\begin{eqnarray}
	\left\langle \Theta_{1/2;i}\right\rangle=\ln\Biggl[\frac{\left(1+2.96
		\sigma_R^2\Omega^{\frac{5}{6}}\right)^2}{\Omega^2\sqrt{\left(1+2.96
			\sigma_R^2\Omega^{\frac{5}{6}}\right)^2+1.2\sigma_R^2\Omega^{\frac{5}{6}}}}\Biggr],
	\label{Eq:Theta}
	\end{eqnarray}
	\begin{eqnarray}
	\left\langle\Delta x_{0;i}^2\right\rangle=\left\langle\Delta y_{0;i}^2\right\rangle=
	0.33\,W_0^2 \sigma_R^2 \Omega^{-\frac{7}{6}},\label{Eq:x02}
	\end{eqnarray}
	\begin{eqnarray}
	\left\langle \Delta \Theta_{1/2;i}^2\right\rangle=\ln\Biggl[1+\frac{1.2\sigma_R^2
		\Omega^{\frac{5}{6}}}{\left(1+2.96\sigma_R^2\Omega^{\frac{5}{6}}\right)^2}\Biggr],
	\label{Eq:Theta2}
	\end{eqnarray}
	\begin{eqnarray}
	\left\langle \Delta \Theta_{1;i}\Delta \Theta_{2;i}\right\rangle=\ln\Biggl[1-\frac{0.8\sigma_R^2
		\Omega^{\frac{5}{6}}}{\left(1+2.96\sigma_R^2\Omega^{\frac{5}{6}}\right)^2}\Biggr],
	\label{Eq:Theta1Theta2}
	\end{eqnarray}
	where $\sigma_R^2=1.23 C_n^2 k^{\frac{7}{6}}L^{\frac{11}{6}}$ is the Rytov parameter, $\Omega{=}{kW_0^2}/{2L}$ is the Fresnel parameter, $k$ is the wavenumber and $L$ is the propagation distance.

	Based on the generated sampling data, the mean fidelity is approximated as
	\begin{eqnarray}
	\bar{F}=\frac{1}{N}\sum\limits_{i=1}^NF\left(\sqrt{\eta_\mathrm{m}\eta\left(\mathbf{v}_i,\chi_i\right)}\right).
	\end{eqnarray}
	Here $\eta_\mathrm{m}$ is the efficiency of constant attenuation.
	The function $\eta\left(\mathbf{v},\chi\right)$ reads
	\begin{eqnarray}
	&\eta\left(\mathbf{v},\chi\right)=\eta_{0}\left(\Theta_1,\Theta_2\right) \label{Tapprox}\\
	&\times\exp\left\{-\left[\frac{r_0/a}
	{R\left(\frac{2}{W_{\rm
				eff}\left(\Theta_1,\Theta_2,\chi\right)}\right)}\right]^{\lambda\bigl(\frac{2}{W_{\rm
				eff}\left(\Theta_1,\Theta_2,\chi\right)}\bigr)}\right\}.\nonumber
	\end{eqnarray}
	where $r_0=\sqrt{x_0^2+y_0^2}$ is the distance between beam and aperture
	centers, $a$ is the radius of the receiver aperture.
	The further parameters introduced in this function are given by
	\begin{eqnarray}
	&W_\textrm{eff}^2\left(\Theta_1,\Theta_2,\chi\right){=}4a^2
	\Bigl[\mathcal{W}\Bigl(\frac{4a^2}{
		W_1\left(\Theta_1\right)W_2\left(\Theta_2\right)}\nonumber\\
	&\times e^
	{\frac{a^2}{W_1^2\left(\Theta_1\right)}\bigl\{1+2\cos^2\!\chi\bigr\}}
	e^
	{\frac{a^2}{W_2^2\left(\Theta_2\right)}\bigl\{1+2\sin^2\!\chi\bigr\}}\Bigl)\Bigr]^{-1},\label{Weff}
	\end{eqnarray}
	\begin{eqnarray}
	&\eta_{0}\left(\Theta_1,\Theta_2\right)\nonumber\\
	&{=}1{-}
	\mathrm{I}_0\Bigl(a^2\Bigl[\frac{1}{W_1^2\left(\Theta_1\right)}{-}\frac{1}{W_2^2\left(\Theta_2\right)}
	\Bigr]\Bigr)
	e^{-a^2\bigl[\frac{1}{W_1^2\left(\Theta_1\right)}{+}
		\frac{1}{W_2^2\left(\Theta_2\right)}\bigr]}\nonumber\\
	&{-}2\left[1{-}e^{-\frac{a^2}{2}\!
		\bigl(\frac{1}{W_1\left(\Theta_1\right)}{-}\frac{1}{W_2\left(\Theta_2\right)}\bigr)^{2}}\!\right]\nonumber\\
	&\times\exp\!\left\{\!{-}
	\Biggl[\!\frac{\frac{(W_1\left(\Theta_1\right)+
			W_2\left(\Theta_2\right))^2}{|W_1^2\left(\Theta_1\right)-W_2^2\left(\Theta_2\right)|}}
	{R\left(\frac{1}{W_1\left(\Theta_1\right)}{-}\frac{1}{W_2\left(\Theta_2\right)}\right)}\!\Biggr]
	^{\!\lambda\left(\!\frac{1}{W_1\left(\Theta_1\right)}{-}\frac{1}{W_2\left(\Theta_2\right)}\right)}\right\},
	\end{eqnarray}
	\begin{eqnarray}
	R\left(\xi\right)=\Bigl[\ln\Bigl(2\frac{1-\exp[-\frac{1}{2} a^2
		\xi^2]}{1-\exp[-a^2\xi^2]\mathrm{I}_0\bigl(a^2\xi^2\bigr)}\Bigr)\Bigr]^{-\frac{1}{
			\lambda(\xi)}},
	\end{eqnarray}
	\begin{eqnarray}
	\lambda\left(\xi\right)&=2a^2\xi^2\frac{e^{-a^2\xi^2}\mathrm{I}_1(a^2\xi^2)}{1-\exp[
		-a^2\xi^2 ] \mathrm{I}_0\bigl(a^2\xi^2\bigr)}\nonumber\\
	&{\times}\Bigl[\ln\Bigl(2\frac{1-\exp[-\frac{1}{2} a^2
		\xi^2]}{1-\exp[-a^2\xi^2]\mathrm{I}_0\bigl(a^2\xi^2\bigr)}\Bigr)\Bigr]^{-1}.
	\end{eqnarray}
	For the considered channel the following parameter values are used:
	\begin{itemize}
		\item Wavelength $\lambda=809$nm;
		\item Initial spot radius $W_{0}=20$mm;
		\item Propagation distance $L=1.6$km;
		\item Deterministic attenuation $\eta_\mathrm{m}=0.7$ $(1.5\mathrm{dB})$;
		\item Aperture radius $a=0.04$m;
		\item Atmospheric index-of-refraction structure constant $C_\mathrm{n}^2=(0.5,\, 1.5,\, 7)\cdot 10^{-14}\;\mathrm{m^{-\frac{2}{3}}}$
	\end{itemize}
	Note that these parameters correspond to conditions of an experimentally implemented free-space experiment in the city of Erlangen \cite{Usenko}. 
	Our model \cite{VSV2016} shows good agreement with the experimental distribution of the transmittance.

\end{document}